\documentstyle[12pt,epsf]{article}

\begin{document}

\begin{titlepage}

\hfill FIAN/TD-23/95

\hfill hep-lat/9601021

\begin{centering}
\vfill

{\bf 
EFFECTIVE POTENTIAL FOR SCALAR FIELD IN THREE  \\
DIMENSIONS: ISING MODEL IN THE FERROMAGNETIC PHASE
}

\vspace{1cm}

M.M.Tsypin\footnote{e-mail: tsypin@lpi.ac.ru}

\vspace{1cm}

{\em Department of Theoretical Physics, \\
P.~N.~Lebedev Physical Institute, Leninsky pr.~53, 117924 Moscow, Russia}

\vspace{2cm}
{\bf Abstract}

\end{centering}

\vspace{0.3cm}\noindent

We compute the effective potential $V_{\rm eff}(\varphi)$ for
one-component real scalar field $\varphi$ in three
Euclidean dimensions (3D) in the case of spontaneously
broken symmetry, from the Monte Carlo simulation of the
3D Ising model in external field at temperatures approaching
the phase transition from below. We study probability
distributions of the order parameter on the lattices from
$30^3$ to $74^3$, at $L/\xi \approx 10$. We find that,
in close analogy with the symmetric case, $\varphi^6$
plays an important role: $V_{\rm eff}(\varphi)$
is very well approximated by the sum of $\varphi^2$,
$\varphi^4$ and $\varphi^6$ terms. An unexpected feature is
the negative sign of the $\varphi^4$ term. As close to the
continuum limit as we can get ($\xi \approx 7.2$), we obtain
\begin{displaymath}
  {\cal L}_{\rm eff}  \approx
  {1 \over 2} \partial_\mu \varphi \partial_\mu \varphi +
  1.7 (\varphi^2 - \eta^2)^2 (\varphi^2 + \eta^2).
\end{displaymath}
We also compute several universal coupling constants and ratios,
including the combination of critical amplitudes 
$C^- (f_1^-)^{-3} B^{-2}$.

\vfill 

\noindent
FIAN/TD-23/95

\noindent
December 1995

\end{titlepage}

\subsection*{Introduction}
This work continues our previous Monte Carlo study \cite{Tsypin},
which was devoted to the effective potential $V_{\rm eff}(\varphi)$
for the theory of one-component real scalar field in three Euclidean
dimensions in the symmetric (paramagnetic, PM) phase. There
we have found that $V_{\rm eff}$ is very well approximated by the
sum of $\varphi^2$, $\varphi^4$ and $\varphi^6$ terms, and
computed universal 4-point and 6-point couplings. Here we turn
to the spontaneously broken (ferromagnetic, FM) phase of the same
theory. To compute the effective potential in the FM case
we use largely the same approach we have developed for the
PM case, with some modifications. As the detailed discussion,
with all necessary references, was given in \cite{Tsypin},
we describe it here only briefly, concentrating mostly on the
points specific for the broken phase.

\subsection*{Monte Carlo computation}
\begin{figure}
\epsfbox{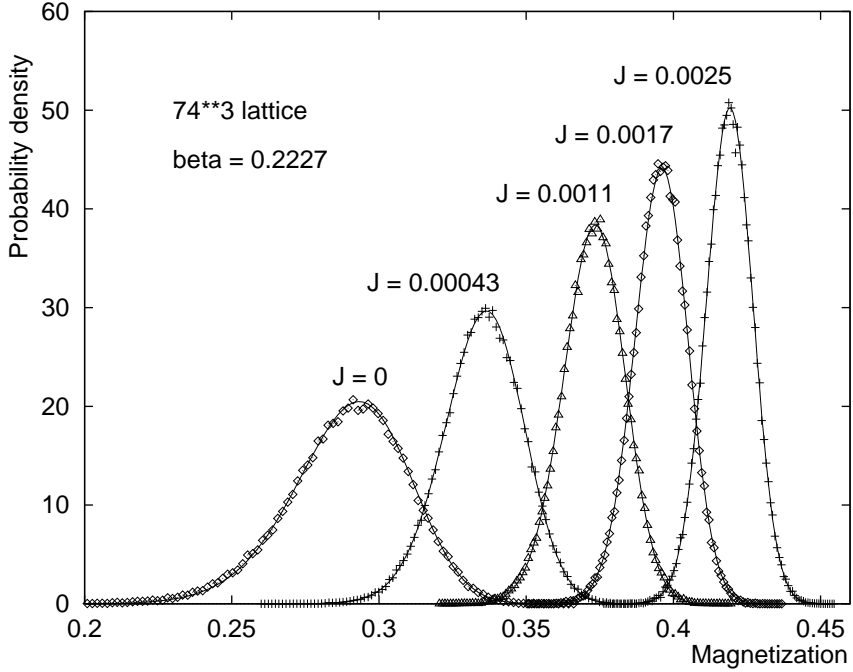}
\caption{\sl
Probability distributions $P(\varphi)$ for magnetization per 
lattice site, for the Ising model (\protect\ref{Ising}). 
The points are from Monte Carlo simulation; the solid line is 
the fit with~(\protect\ref{preexp}),
$V_{\rm eff}(\varphi) = r \varphi^2 + u \varphi^4 + w \varphi^6$
(all 5 histograms are fitted simultaneously with one $V_{\rm eff}$).
 }
\label{fig246}
\end{figure}

We consider the Ising model in external field,
\begin{equation} \label{Ising}
  {\cal Z} = \sum_{ \{\phi_i \} } \exp \Bigl\{
  \beta \sum_{<ij>} \phi_i \phi_j + J \sum_i \phi_i \Bigr\},
  \qquad \phi_i = \pm 1,
\end{equation}
on a simple cubic lattice of the size $L^3$ (from $30^3$ to $74^3$) 
with periodic boundary conditions. The critical coupling is 
$\beta_c \approx 0.221655$ \cite{Blote95}. We use Swendsen-Wang
cluster Monte Carlo algorithm \cite{SW} to generate the Boltzmann
ensemble of configurations, for given coupling $\beta$ and
external field $J$. For every configuration we measure the
order parameter (magnetization per site)
\begin{equation}
  \varphi = {1 \over N} \sum_{i=1}^N \phi_i,
\end{equation}
where $N=L^3=\Omega$ is the total number of sites. Thus we obtain
probability distributions $P(\varphi)$, in form of histograms
(Fig. \ref{fig246}). The relation between $P(\varphi)$ and
$V_{\rm eff}(\varphi)$, for sufficiently large volume $\Omega$, is
\cite{Tsypin}
\begin{equation} \label{preexp}
  P(\varphi) \propto
  \sqrt{ V''_{\rm eff}(\varphi) }
  \exp \left\{  - \Omega V_{\rm eff}(\varphi) + \Omega J \varphi \right\}.
\end{equation}
Now one can check whether it is possible to fit the set of
probability distributions for given $\beta$ and several values
of $J$ with this formula, using this or that ansatz for 
$V_{\rm eff}(\varphi)$.

\begin{figure}
\epsfbox{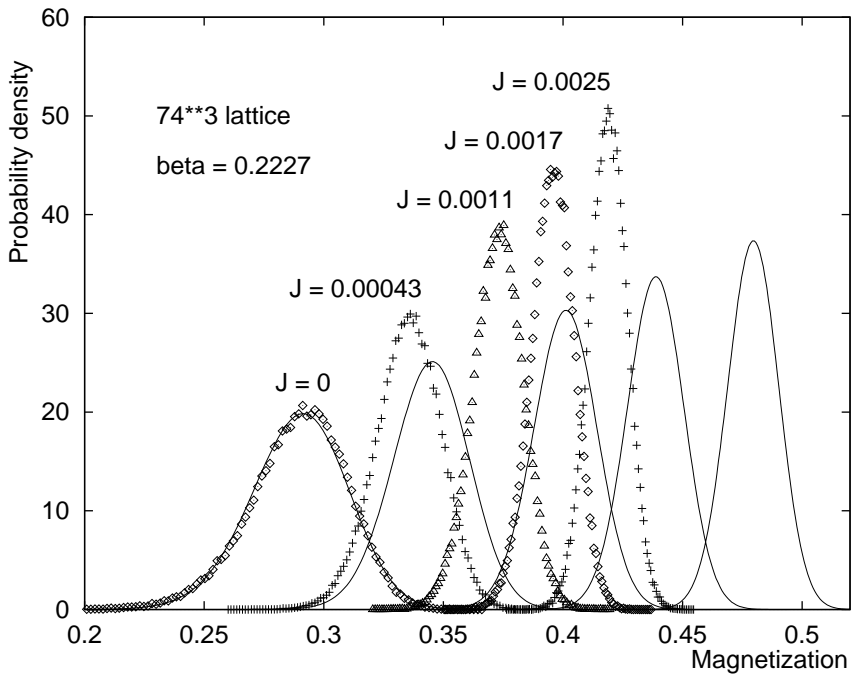}
\caption{\sl
The Monte Carlo points are the same as at Fig.~\protect\ref{fig246}.
The solid line is given by (\protect\ref{preexp}),
$V_{\rm eff}(\varphi) = r \varphi^2 + u \varphi^4$,
where $r$ and $u$ are fixed by the requirement that spontaneous
magnetization and zero field susceptibility are correctly 
reproduced.
 }
\label{fig24}
\end{figure}

We see that the situation is analogous to that in the PM phase
\cite{Tsypin}: the ansatz $V_{\rm eff}(\varphi) = r \varphi^2 + u \varphi^4$
works poorly, while
\begin{equation} 
V_{\rm eff}(\varphi) = r \varphi^2 + u \varphi^4 + w \varphi^6
\end{equation}
provides a perfect fit (Figs. \ref{fig246}, \ref{fig24}).
In the remaining part of the paper we discuss extraction of
universal quantities from $r$, $u$, $w$ obtained from such fits.

\subsection*{Problems with fitting procedures}
In the FM phase several complications arise, that make determination
of parameters of the effective potential much more tricky than in
the PM phase.

First, it turns out that the ratio of lattice size to correlation
length $L/\xi$ has to be much larger in the FM phase than in the
PM phase, to get finite size effects under control. The formula
(\ref{preexp}) is valid in the limit of large $L/\xi$. It turns
out that to get finite volume corrections to $V_{\rm eff}$ as small
as they are in the PM phase at $L/\xi=4$, one has to go to $L/\xi$
as large as 10 in the FM phase.

Thus it is difficult to go to large $\xi$, as the lattice size becomes
prohibitively large. This, in its turn, makes it very difficult to
study the finite cutoff effects and to extrapolate to $\xi \to \infty$.
While it was possible to do such an extrapolation, with a reasonable 
degree of confidence, in the PM case, where we could study $\xi$
as large as 14.1, in the FM case we can only get a qualitative
estimate of finite cutoff corrections, and have no choice other
than to take the values from our largest lattice, without 
attempting any extrapolation.

Secondly, the applicability of eq. (\ref{preexp}) for various values
of $\varphi$ depends on the ratio $L/\xi(\varphi)$, where
$\xi(\varphi)$ is determined by $V''_{\rm eff}(\varphi)$, and
deteriorates quickly at $|\varphi| < M$, where $M$ is spontaneous
magnetization. Thus we have to use for the fitting only the
subset of data corresponding to $|\varphi| > M$, losing
considerable amount of information (practically, a half of the
$J=0$ histogram).

So one has to take into account that

1) Including larger values of $J$ into the fit reduces 
statistical errors, but increases systematic errors connected
with the finite cutoff (i.e. the smallness of $\xi$), as 
larger $J$ mean smaller $\xi$.

2) It would be nice to use the information contained in the
whole $J=0$ histogram, including $|\varphi| < M$.

One can see how this works, from Table \ref{T1}, where we have 
compiled some observables that can be measured directly, as
well as position $M$ of the minimum of $V_{\rm eff}$ (that corresponds
to spontaneous magnetization $\langle|\varphi|\rangle$) and its
second derivative in the minimum $V''_{\rm eff}(M)$ (that corresponds
to inverse susceptibility $\chi^{-1}$), obtained by fitting
either 2 or 3 histograms. 

From Table \ref{T1} one observes that

1) The best results for both $M$ and $V''_{\rm eff}(M)$, as far as 
statistical errors are concerned, come from the direct 
measurement (apparently because the region $|\varphi| < M$
is lost for the fitting).

2) Fitting 2 histograms leads to results compatible
with direct measurement, for all lattices, but with
unsatisfactorily large statistical errors.

3) Fitting 3 histograms leads to considerable reduction
of statistical errors. However, with the exception of
the two largest lattices, there is a serious deviation
of $M$ and $V''_{\rm eff}(M)$ from the direct measurement.
This is a manifestation of a systematic error
associated with the finite cutoff, due to the
smallness of the correlation length at $J=J_2$
(corresponding to the third histogram) on our smaller
lattices.

\subsection*{The improved fitting procedure}
Thus both fitting procedures in the Table \ref{T1}
are far from satisfactory. So we have designed the
following procedure that uses the $J=0$ data to
significantly reduce statistical errors of the
2-histogram fit: one should fix the minimum of
$V_{\rm eff}$ at the value of $M$ obtained from the
direct measurement. This leads to 2-parameter fit instead
of the 3-parameter one, with corresponding reduction of
statistical errors. The results are collected in Table \ref{T2}.

One observes that: 1) the quality of fits, as indicated by
$\chi^2$, is very high: there is no discrepancy between
Monte Carlo histograms and the fit, other than statistical noise;
2) $V''_{\rm eff}(M)$ is completely consistent with the direct 
measurement; 3) Fitting 5 histograms for our largest lattice
without fixing the position of the minimum of $V_{\rm eff}$
(Fig. \ref{fig246} and the last column in Table \ref{T2})
provides both $M$ and $V''_{\rm eff}(M)$ completely consistent
with the direct measurement. All parameters are consistent
with the 2-histogram fit.

\subsection*{Extraction of universal parameters}
In the FM case the extraction of universal, i.e. dimensionless,
parameters of $V_{\rm eff}$ from the data becomes less trivial than
in the PM case.

Having measured $r$, $w$, $u$ from the fits and the field 
renormalization constant $Z$ from the small-momentum behavior of 
the propagator,  we obtain the effective Lagrangian in the form
\begin{equation}
  {\cal L}_{\rm eff} =
  {1 \over 2} Z^{-1} \partial_\mu \varphi \partial_\mu \varphi +
  r \varphi^2 + u \varphi^4 + w \varphi^6 .
\end{equation}
Then we change the scale of $\varphi$, introducing the renormalized 
field $\varphi_R = Z^{-1/2} \varphi$:
\begin{equation}
  {\cal L}_{\rm eff} = 
  {1 \over 2} \partial_\mu \varphi_R \partial_\mu \varphi_R +
  Zr \varphi_R^2 + Z^2 u \varphi_R^4 + Z^3 w \varphi_R^6 .
\end{equation}
The coefficients in front of $\varphi_R^2$, $\varphi_R^4$ and
$\varphi_R^6$ have, correspondingly, dimensionalities $m^2$, $m$ and 1.
In the PM case it was natural to choose $\sqrt{2Zr}$
as a scale factor, and to use it to render the 4-point coupling
dimensionless, obtaining two dimensionless
parameters: $g_4 = Z^2u/\sqrt{2Zr}$ and $g_6 = Z^3w$.

In the FM case the coefficient in front of $\varphi_R^2$ does not
determine the correlation length any more. Moreover, $r$ is 
determined with very large statistical errors. So we have to
find something different that could serve as a scale factor.
Spontaneous magnetization seems to be the best choice:
$\langle|\varphi_R|\rangle^2 \equiv M_R^2 = Z^{-1}M^2 $
has the dimensionality of mass and very small statistical error.

\begin{figure}
\epsfbox{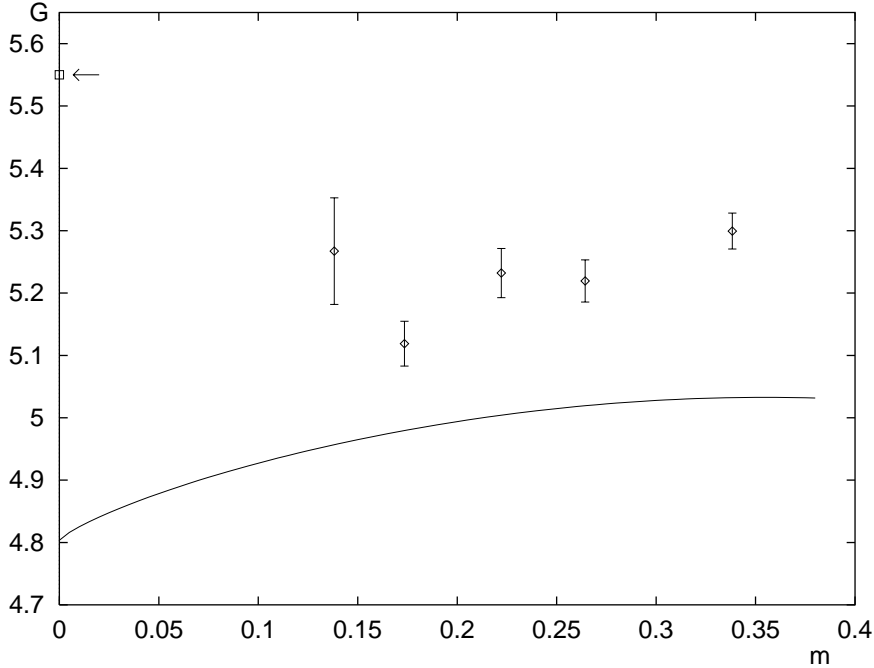}
\caption{\sl
Dimensionless parameter 
$G = Z^{3/2} \protect\sqrt{V''} M^{-2} = \chi \xi^{-3} M^{-2}$,
as a function of inverse correlation length. The diamonds are
our Monte Carlo results, the square is from \protect\cite{BTW95},
and the solid line is from approximants given by Liu and Fisher
\protect\cite{LF89} (see Appendix). The line has an uncertainty
of order 3\% in the overall factor. If one combines our points 
with $m$-dependence from Liu and Fisher's approximants, one gets
in the scaling limit $ G \to C^- (f_1^-)^{-3} B^{-2} \approx 5.0\,$.
 }
\label{figG}
\end{figure}

Thus the coefficient in front of $\varphi_R^4$ provides a 
dimensionless parameter $Z^3u/M^2$ (included in Table \ref{T2}).
Another interesting dimensionless parameter is obtained from 
the mass:
\begin{equation} 
  G \equiv m/M_R^2 = { Z^{3/2} \sqrt{V''} \over M^2 } .
\end{equation}
Its remarkable property is that, being a special universal
combination of critical amplitudes, it can be measured,
to high precision, without any fitting (Fig. \ref{figG}):
\begin{eqnarray} 
  G & = & \chi \xi^{-3} M^{-2},  \label{G}  \\
  \lim_{\xi \to \infty} G & = & C^- (f_1^-)^{-3} B^{-2}. 
\end{eqnarray}
(See Appendix for notation).
More dimensionless parameters are obtained from the third and fourth
derivatives of $V_{\rm eff}$ in the minimum (Table \ref{T2}), such as 
\begin{equation} \label{kappa}
  \kappa = { V''' M \over 6 V'' }.
\end{equation}

\subsection*{Expansion around the minimum}
Consider the situation in the vicinity of the minimum of $V_{\rm eff}$.
One can write $\varphi(x) = M + \tilde\varphi(x)$, where
$\tilde\varphi$ is a small deviation. Then for $\tilde\varphi$
we get
\begin{equation}
  {\cal L}_{\rm eff} = {1 \over 2} Z^{-1} \partial_\mu \tilde\varphi
     \partial_\mu \tilde\varphi + 
     {1 \over 2} V''\tilde\varphi^2 +
     {1 \over 3!} V'''\tilde\varphi^3 + 
     {1 \over 4!} V''''\tilde\varphi^4 + \ldots\,,
\end{equation}
where all derivatives of $V_{\rm eff}$ are taken in its minimum.
After changing the scale, $\tilde\varphi = \sqrt{Z} \tilde\varphi_R$,
\begin{equation}
  {\cal L}_{\rm eff} = {1 \over 2} \partial_\mu \tilde\varphi_R
     \partial_\mu \tilde\varphi_R + 
     {1 \over 2} Z V''\tilde\varphi_R^2 +
     {1 \over 3!} Z^{3/2} V'''\tilde\varphi_R^3 + 
     {1 \over 4!} Z^2 V''''\tilde\varphi_R^4 + \ldots\,,
\end{equation}
Now we obtain the mass,
\begin{equation}
  m = \xi^{-1} = \sqrt{ZV''}
\end{equation}

\begin{figure}
\epsfbox{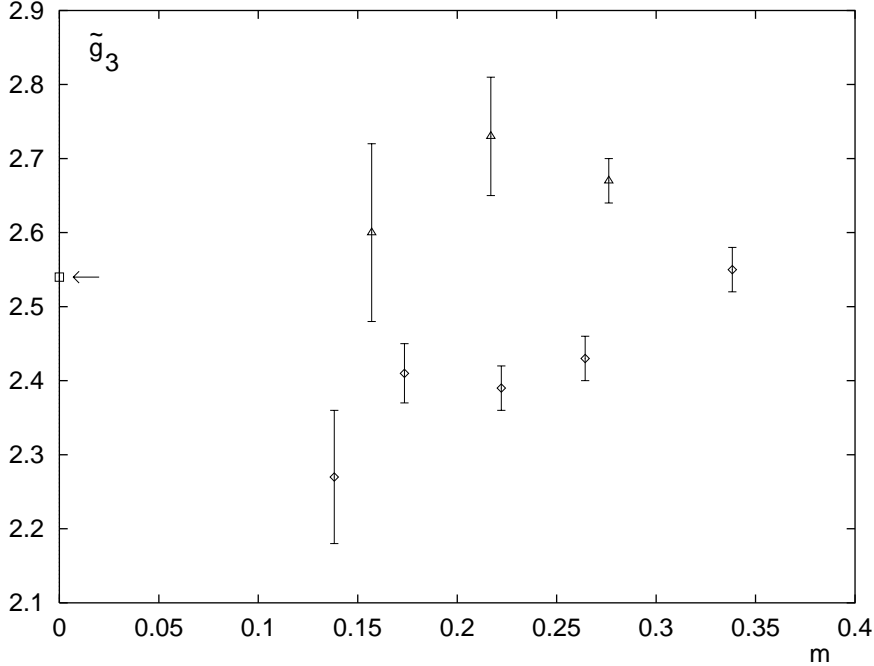}
\caption{\sl
Dimensionless parameter $\tilde g_3$, defined by eq. (\protect\ref{g3}),
as a function of inverse cor\-relation length. Diamonds are ours,
triangles are from \protect\cite{KiPa93}, and the square is from
\protect\cite{BTW95}.
 }
\label{fig-g3}
\end{figure}

\noindent
and dimensionless 3-point and 4-point couplings:
\begin{eqnarray}
  \tilde g_3 & = & {1 \over 3!} { Z^{3/2} V''' \over m^{3/2} } =
      {1 \over 6} \Bigl( Z^{-1} V'' \Bigr)^{-3/4} V'''   \label{g3} \\
  \tilde g_4 & = & {1 \over 4!} {Z^2 V'''' \over m} =
      {1 \over 24} {Z^{3/2} V'''' \over \sqrt{V''} }.
\end{eqnarray}
They are in the following correspondence with $\tilde g_R^{(3)}$ and
$\tilde g_R^{(4)}$ computed by J.-K.~Kim and A.~Patrascioiu \cite{KiPa93}:
\begin{eqnarray}
  \tilde g_3 & = & {1 \over 6} \tilde g_R^{(3)},  \\
  \tilde g_4 & = & {1 \over 24} \Bigl[ 
         \tilde g_R^{(4)} +  3 ( \tilde g_R^{(3)} )^2 \Bigr].
\end{eqnarray}
As can be seen from Fig. \ref{fig-g3}, our values of $\tilde g_3$
are somewhat lower than than those obtained in \cite{KiPa93} by
direct measurement of the 3-point correlation function.
As a consistency check, we have computed $\tilde g_3$ and 
$\tilde g_4$ for two largest lattices directly from 1-particle
irreducible 3- and 4-point functions, and obtained
\begin{equation}
\begin{array}{llll}
  58^3, & \beta=0.2232: & \tilde g_3 = 2.41(7), & \tilde g_4 = 5.4(7); \\
  74^3, & \beta=0.2227: & \tilde g_3 = 2.35(8), & \tilde g_4 = 5.1 \pm 2.1,
\end{array}
\end{equation}
in complete agreement with Table \ref{T2}.

\subsection*{Compact formula for $V_{\rm eff}$}
We see that quite a few dimensionless parameters of $V_{\rm eff}$
can be constructed. However, one would like to have $V_{\rm eff}$
written down in a concise form, so that its shape can be better
understood. For the PM phase it could be written \cite{Tsypin} as 
\begin{equation}
  V_{\rm eff} = {1 \over 2} m^2 \varphi_R^2 + m g_4 \varphi_R^4 +
            g_6 \varphi_R^6,
\end{equation}
with the scale parameter $m$ and two dimensionless (and thus universal)
parameters $g_4$ and $g_6$.

In the broken phase one can take as a starting point the following 
observation. If the constant term in $V_{\rm eff}$ is chosen in such
a way that $V_{\rm eff}=0$ in the minimum, it has double zeros at
$\varphi_R = \pm M_R$, and must be proportional to 
$(\varphi_R^2 - M_R^2)^2$. Thus it must take the form
\begin{equation}
  V_{\rm eff}(\varphi_R) = (\varphi_R^2 - M_R^2)^2 ( a \varphi_R^2 + b M_R^2),
\end{equation}
with two dimensionless parameters $a$ and $b$, that can be expressed
via $G$ and $\kappa$ (\ref{G},\ref{kappa}), leading to
\begin{equation}
  V_{\rm eff}(\varphi_R) = 
    {1 \over 16} G^2 (\varphi_R^2 - M_R^2)^2 
    \Bigl[ 2 M_R^2 + (2\kappa - 1)(\varphi_R^2 - M_R^2) \Bigr].
\end{equation}
Unlike the symmetric phase, where we had the range of $\xi$ sufficient
to discuss extrapolation to $\xi \to \infty$, our results for the
broken phase indicate the existence of the finite cutoff effects,
but are clearly not sufficient to extrapolate to $\xi \to \infty$.
So we can only write down the result for our largest $\xi$
(the last columns of Tables \ref{T1} and \ref{T2}):
\begin{eqnarray}
  G & = & 5.27(9), \qquad \kappa = 0.998(14), \\
  V_{\rm eff} & = & 1.73(6) \cdot (\varphi_R^2 - M_R^2)^2 
    \Bigl[ 2 M_R^2 + 1.00(3) \cdot (\varphi_R^2 - M_R^2) \Bigr].
      \label{final}
\end{eqnarray}
The hope is that these numbers are already close to the continuum
limit. It is interesting to observe that $V_{\rm eff}$ turns out to be
proportional to $(\varphi^2 - M^2)^2 (\varphi^2 + M^2)$. This
sheds some light on the negative sign of $\varphi^4$, as
$(\varphi^2 - 1)^2 (\varphi^2 + 1) = 1 - \varphi^2 - \varphi^4 +
\varphi^6$.

\subsection*{Comparison with analytical results}

\begin{figure}
\epsfbox{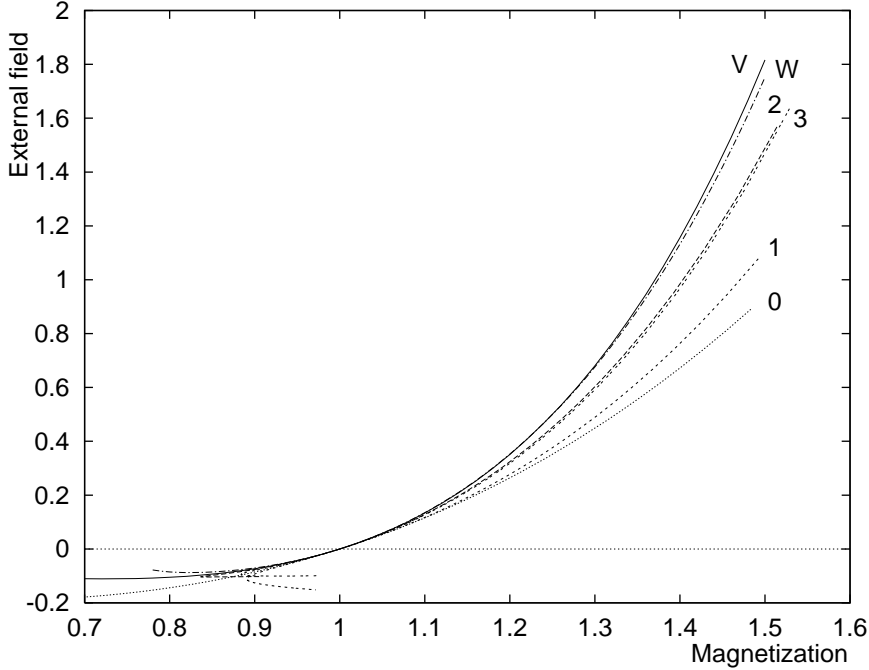}
\caption{\sl
The plots of the ``Ising equation of state'', 
$J(\varphi) = dV_{\rm eff}(\varphi)/d\varphi$, with effective potentials
computed by different means, all of them normalized in such a way
that the minimum of $V_{\rm eff}$ is at $M=1$, and $V''_{\rm eff} = 1$.
The curves 0--3 are from the $\varepsilon$-expansion of the parametric
representation of the equation of state \protect\cite{Zi89},
in orders $\varepsilon^0 \ldots \varepsilon^3$, respectively.
The curve $W$ is from \protect\cite{BTW95}, and the curve $V$ 
represents our result (\protect\ref{final}), which reduces in this
normalization to $J={1 \over 8} (-\varphi - 2\varphi^3 + 3\varphi^5)$.
 }
\label{fig-eqst}
\end{figure}

The only analytical results on the effective potential in the
broken phase we know of are from the $\varepsilon$-expansion
of the scaling equation of state (see \cite{Zi89}), and those
obtained recently by Berges, Tetradis and Wetterich \cite{BTW95}
using a method based on an exact flow equation for a coarse
grained free energy. In their notation,
\begin{equation}
  U_R(\rho_R) \equiv V_{\rm eff}(\varphi_R), 
    \qquad \rho_R \equiv {1 \over 2} \varphi_R^2,
\end{equation}
they obtain
\begin{eqnarray}
  { \hat \lambda_R \over \rho_{0R} } & \equiv &
  { U''_R(\rho_{0R}) \over \rho_{0R} } = 61.6, \\[5pt]
  \hat \nu_R & \equiv & U'''_R(\rho_{0R}) = 107.
\end{eqnarray}
This translates, in our notation, to
\begin{eqnarray}
  G & = & \biggl( {1 \over 2} {\hat\lambda_R \over \rho_{0R} }
      \biggr)^{1/2} \approx 5.55 \ ,  \\
  \kappa  & \approx & 1.08 \ ,   \\
  \tilde g_3 & \approx & 2.54 \ .
\end{eqnarray}
These values are included in Figs. \ref{figG}, \ref{fig-g3} and are 
5--8 \%
higher than our Monte Carlo results. Berges, Tetradis and Wetterich
give also a complicated formula that serves as an approximant for
$J(\varphi) = V'_{\rm eff}(\varphi)$. The corresponding curve is
shown in Fig. \ref{fig-eqst}, and goes quite close to ours.
Also included are three known terms of the $\varepsilon$-expansion
of the parametric representation of equation of state \cite{Zi89}.

\subsection*{Conclusions}
Here are our main conclusions about the universal properties of
the effective potential for the broken phase of the 3D Ising model
in the scaling region:

(1) The form $r \varphi^2 + u \varphi^4 + w \varphi^6$ provides
a very good approximation for $V_{\rm eff}(\varphi)$ (Fig.~\ref{fig246}),
while without the $\varphi^6$ term this is not achievable
(Fig. \ref{fig24}) --- the same result as in the symmetric phase
\cite{Tsypin}.

(2) Not only $r$, but also $u$ turns out to be negative 
(Table \ref{T2}).

(3) Quantitative results can be summarized by eq. (\ref{final}).

(4) The combination of critical amplitudes 
$G = C^- (f_1^-)^{-3} B^{-2}$ (Fig. \ref{figG}) seems to be 
a quantity very suitable for a precise Monte Carlo computation,
especially using improved estimators, as it does not suffer
from uncertainties in $T_c$ and critical exponents.

It should be noted that, thinking about $V_{\rm eff}$ in terms of 
Taylor expansion around the minimum, one would need at least 
4 parameters to describe our data
($M$, $V''_{\rm eff}(M)$, $V'''_{\rm eff}(M)$ and $V''''_{\rm eff}(M)$),
and the symmetry $\varphi \to - \varphi$ would be lost.
Thus the possibility to approximate $V_{\rm eff}$ with 
$r \varphi^2 + u \varphi^4 + w \varphi^6$, that has only
3 parameters and respects the symmetry, is by no means trivial,
and demonstrates, once more, a special role of $\varphi^6$
in three dimensions.

\subsection*{Acknowledgements}
It is a pleasure to thank Alexander von Humboldt Foundation
for the fellowship and support, and the Institute of Theoretical
Physics E of RWTH Aachen, where this work has been initiated,
for the kind hospitality. This work was partially supported
by a fellowship of INTAS Grant 93-2492 within the research program
of International Center for Fundamental Physics in Moscow.

\subsection*{Appendix}
In Table \ref{T1} and Fig. \ref{figG} we have included, for
comparison, the values obtained from approximants given by
Liu and Fisher \cite{LF89}:
\begin{equation}
\begin{array}{lll}
  t = (T-T_c)/T_c = (\beta_c - \beta)/ \beta ,  \\[5pt]
  \xi^2(t) = (f_1^-)^2|t|^{-2\nu} - 0.430, & f_1^- = 0.2502(8), 
     & 2\nu = 1.267,  \\[5pt]
  \chi(t) = C^-|t|^{-\gamma} - 2.71, & C^- = 0.220(4), 
     & \gamma = 1.2395,  \\[2pt]
  M(t) = B |t|^\beta (1 - 0.256 \sqrt{|t|}), & B = 1.71(2),
     & \beta = 0.3305 \,.
\end{array} \label{LF}
\end{equation}
We use the best available estimate of the critical coupling,
$\beta_c = 0.221655$ \cite{Blote95}, rather than the value
0.22163 originally associated with these approximants.

\newpage

\def\refname


\begin{table}
{\small
 \begin{center}
  \begin{tabular}{llllll}
   \hline\rule{0pt}{12pt}
    $L^3$   & $30^3$  & $38^3$  & $46^3$  & $58^3$  & $74^3$   \\
   \hline\rule{0pt}{12pt}
    $\beta$ & 0.226   & 0.2246  & 0.2239  & 0.2232  & 0.2227   \\
    $J_0$   & 0.0     & 0.0     & 0.0     & 0.0     & 0.0      \\
    $J_1$   & 0.004   & 0.0022  & 0.0014  & 0.00078 & 0.00043  \\
    $J_2$   & 0.010   & 0.0056  & 0.0035  & 0.0020  & 0.0011   \\
    $J_3$   &         &         &         &         & 0.0017   \\
    $J_4$   &         &         &         &         & 0.0025   \\
    $N_{\rm config}$ 
        & $3 \times 360000$
        & $3 \times 360000$
        & $3 \times 360000$
        & $3 \times 360000$
        & $5 \times 100000$  \\
   \hline\rule{0pt}{12pt}
    $Z^{-1}_{(J=0)}$ 
        & 0.3167(11) 
        & 0.3166(13) 
        & 0.3149(15) 
        & 0.3174(14) 
        & 0.3133(32)  \\
    $M=\langle|\varphi|\rangle_{J=0}$
        & 0.44975(17)
        & 0.39972(19)
        & 0.36743(18)
        & 0.32700(16)
        & 0.2892(3)   \\
    $M$ from (\ref{LF})
        & 0.4468 & 0.3962 & 0.3640 & 0.3235 & 0.2856  \\
    $V''$ from 
        & 0.0365(1) & 0.02207(8) & 0.01558(7) & 0.00958(4) & 0.00597(6) \\
    the propagator \\
    $\chi^{-1}$ from (\ref{LF})
        & 0.0374 & 0.0224 & 0.0158 & 0.00982 & 0.00600 \\
    $\xi = (Z^{-1}/V'')^{1/2}$
        & 2.946(6) & 3.788(10) & 4.496(14) & 5.756(17) & 7.24(5) \\
    $G \equiv Z^{3/2}\sqrt{V''}M^{-2}$
        & 5.30(3) & 5.22(3) & 5.23(4) & 5.12(4) & 5.27(9) \\
   \hline\rule{0pt}{12pt}
    & \multicolumn{5}{c}{Fitting 2 histograms ($J = J_0$, $J_1$) at 
                       $|\varphi| > M$}  \\[5pt]
    $\sum \chi^2$  & 102 & 106 & 111 &  92 &  83  \\
    $N_{\rm bins}$ & 104 &  98 & 108 & 109 & 106  \\
    $M$   & 0.4491(8)  & 0.3993(13) & 0.3672(7) & 0.3271(7) & 0.2899(7) \\
    $V''$ & 0.0353(13) & 0.0217(12) & 0.0154(5) & 0.0096(3) & 0.0063(3) \\
   \hline\rule{0pt}{12pt}
    & \multicolumn{5}{c}{Fitting 3 histograms ($J = J_0$, $J_1$, $J_2$) at 
                       $|\varphi| > M$}  \\[5pt]
    $\sum \chi^2$  & 188 & 187 & 194 & 209 & 109  \\
    $N_{\rm bins}$ & 178 & 175 & 187 & 188 & 112  \\
    $M$   & 0.4470(4)  & 0.3979(4) & 0.3666(9) & 0.3281(5) & 0.2894(9) \\
    $V''$ & 0.0320(5)  & 0.0204(6) & 0.0149(3) & 0.0102(3) & 0.0060(3) \\
   \hline
  \end{tabular}
 \end{center}
 }
\caption{
Monte Carlo numerical results, computed by
direct measurement of some observables, as well as
by two different fitting procedures. Values of spontaneous
magnetization and inverse susceptibility from approximants
proposed by Liu and Fisher (see Appendix) are listed for 
comparison. $N_{\rm config}$ is the number of configurations used.
$V'' \equiv V''_{\rm eff}(M)$.
$\sum \chi^2$ is the sum of $\chi^2$ for all histograms
used in the fit (it is minimized by the fit). $N_{\rm bins}$ is the
total number of bins in these histograms. The numbers in
parentheses are standard deviations of the last decimal
digits.} \label{T1}
\end{table}

\tabcolsep=0.2em
\begin{table}
{\small
 \begin{center}
  \begin{tabular}{lllllll}
   \hline\rule{0pt}{12pt}
    $L^3$   & $30^3$  & $38^3$  & $46^3$  & $58^3$  & $74^3$ 
      & \multicolumn{1}{|l}{$74^3$}  \\
   \hline\rule{0pt}{12pt}
    & \multicolumn{5}{c|}{Fitting 2 histograms ($J = J_0$, $J_1$) at 
                       $|\varphi| > M$,} 
      & \multicolumn{1}{c}{Fitting 5} \\
    & \multicolumn{5}{c|}
      {keeping minimum of $V_{\rm eff}(\varphi)$ fixed at $\varphi = M$}
      & \multicolumn{1}{c}{histograms} \\
    & \multicolumn{5}{c|}{} 
      & \multicolumn{1}{c}{at $|\varphi| > M$} \\[5pt]
   \hline
    $\sum \chi^2$ \rule{0pt}{12pt}  & 103 & 106 & 111 &  92 &  85 & 200 \\
    $N_{\rm bins}$ & 104 &  98 & 108 & 109 & 106 & 204 \\
    $r$  
         & -0.00077(21) 
         & -0.00084(13)
         & -0.00071(8)
         & -0.00036(8)
         & -0.00040(10)
         & -0.00038(4)  \\
    $u$
         & -0.0186(9)
         & -0.0120(7)
         & -0.0091(5)
         & -0.0078(6)
         & -0.0042(10)
         & -0.0044(3)  \\
    $w$
         & 0.0676(12)
         & 0.0612(13)
         & 0.0581(11)
         & 0.059(2)
         & 0.052(3)
         & 0.0532(7)  \\
    $M$   
         & 0.44975
         & 0.3997
         & 0.3674
         & 0.3270
         & 0.2892
         & 0.2893(6)  \\
    $V''$ 
         & 0.03625(26)
         & 0.02211(15)
         & 0.01555(9)
         & 0.00955(7)
         & 0.00599(12)
         & 0.00599(13) \\
    $m=\sqrt{ZV''}$
         & 0.3383(12)
         & 0.2643(9)
         & 0.2222(7)
         & 0.1735(7)
         & 0.1382(14)
         & 0.1382(15)  \\
    $V'''$
         & 0.537(4)
         & 0.353(3)
         & 0.265(2)
         & 0.1870(23)
         & 0.123(3)
         & 0.1239(10)  \\
    $V''''$
         & 4.48(7)
         & 3.23(6)
         & 2.60(4)
         & 2.09(5)
         & 1.47(8)
         & 1.497(10) \\[3pt]
    & \multicolumn{6}{c}{The following parameters are dimensionless:} \\[5pt]
    $Z^2u/M^2$
         & -2.90(14)
         & -2.37(14)
         & -2.17(12)
         & -2.29(19)
         & -1.6(4)
         & -1.72(11) \\
    $g_6=Z^3w$
         & 2.13(4)
         & 1.93(4)
         & 1.86(4)
         & 1.85(5)
         & 1.70(10)
         & 1.73(2) \\
    $\tilde g_3$
         & 2.55(3)
         & 2.43(3)
         & 2.39(3)
         & 2.41(4)
         & 2.27(9)
         & 2.29(2) \\
    $\tilde g_4$
         & 5.50(10)
         & 5.08(11)
         & 4.92(9)
         & 4.99(14)
         & 4.5(3)
         & 4.60(8) \\[3pt]
    & \multicolumn{6}{c}{The following parameters are dimensionless
         and do not contain $Z$:} \\[5pt]
    $\kappa = V'''M/6V''$
         & 1.110(15)
         & 1.065(16)
         & 1.045(13)
         & 1.067(20)
         & 0.99(4)
         & 0.998(14) \\
    $V''''M^2/24V''$
         & 1.041(23)
         & 0.972(24)
         & 0.942(20)
         & 0.975(30)
         & 0.86(7)
         & 0.87(2) \\
    $\tilde g_4 / \tilde g_3^2$
         & 0.844(5)
         & 0.857(5)
         & 0.863(4)
         & 0.857(6)
         & 0.878(11)
         & 0.875(4) \\[3pt]
   \hline
  \end{tabular}
 \end{center}
  }
\caption{
Numerical results obtained by the fitting procedure
that we find most successful. For the largest lattice,
fitting all 5 histograms gives the same results (the last
column), providing an additional consistency check.
All derivatives of $V_{\rm eff}$ are taken in the minimum.} \label{T2}
\end{table}

\end{document}